\documentclass[aps,pre,twocolumn,superscriptaddress,longbibliography]{revtex4-2}
\usepackage[colorlinks=true,linkcolor=blue,citecolor=blue]{hyperref}
\usepackage{amsmath,amssymb}
\usepackage[utf8]{inputenc}
\usepackage[T1]{fontenc}
\usepackage{url}
\usepackage{adjustbox}
\usepackage{subcaption}
\usepackage{graphicx}
\usepackage{epstopdf}
\usepackage{bm}
\usepackage{bbm}
\usepackage{xcolor}
\usepackage{color}
\usepackage{ulem}
\usepackage[toc]{appendix}
\usepackage{subfloat}
\usepackage{ragged2e}
\usepackage{caption}
\usepackage{braket}
\DeclareCaptionFormat{myformat}{\justifying#1#2#3}
\begin{document}
 
\title{Study of entanglement and phase transitions in the coupled top systems with standard and nonstandard symmetries}

\author{Rashmi Jangid Sharma}
\email{jangid.rashmi@gmail.com}
\affiliation{Department of Applied Sciences and Humanities, B. K. Birla Institute of Engineering and Technology, Pilani 333031, India} 

\author{Jayendra N. Bandyopadhyay}
\email{jnbandyo@gmail.com}
\affiliation{Department of Physics, Birla Institute of Technology and Science Pilani, Pilani Campus, Vidya Vihar, Pilani, Rajasthan 333031, India.}

\begin{abstract}
 
We study classical and quantum versions of a coupled top system in the absence and the presence of nonlinear torsion in the individual top. The model without the torsion and couples two identical tops is well-known in the literature as the Feingold-Peres (FP) model. The permutation and chiral symmetries are preserved in the FP model. This model is classified under the BDI or chiral orthogonal symmetry class, one of the recently proposed nonstandard symmetry classes. For the nonzero torsional cases, we study two different models: (i) identical torsional term in the individual top (NZT-I model); (ii) non-identical torsional term due to their opposite sign in the individual top (NZT-II model). The NZT-I model has the permutation symmetry, but no chiral symmetry; hence, this model is classified under the standard three-fold symmetry classes. On the other hand, the NZT-II model does not have permutation symmetry but has chiral symmetry; hence, this model is also classified as a nonstandard BDI symmetry class. In this study, we investigate the role of underlying symmetries on the entanglement between the two tops. Moreover, we explore the interrelations among classical phase space dynamics, energy transitions, and the entanglement between the tops.

\end{abstract}

\maketitle

\section{Introduction}
\label{sec1}

Kicked top and coupled kicked top are well-known periodically delta-kicked systems which have been studied in the context of classical and quantum chaos and quantum entanglement \cite{haake1987classical,kus1987,Haake,Miller, JNBandArul01,JNBandArul,demkowicz2004,herrmann2020,tanaka2002saturation, fujisaki2003dynamical,sharma2018}. These kicked systems have also been experimentally realized in cold atoms and nuclear magnetic resonance (NMR) \cite{chaudhury2009quantum,ghose2008chaos,mahesh2019}. In the cold atoms setup, the delta-kicks are controlled by applying periodic laser pulses. On the other hand, in the NMR setup, these kicks are realized using radio-frequency pulses. In principle, one can set the time interval between the consecutive laser and radio-frequency pulses to be so small that the corresponding frequency becomes so large that it does not resonate with any internal transitions. At this high-frequency limit, one can realize an effective time-independent Hamiltonian. Theoretically, one can derive this effective Hamiltonian using Floquet theory based perturbation theory \cite{Rahav,GoldmanandDalibard,JNBandTGS}. 

The standard kicked top model consists of linear rotation and nonlinear torsion \cite{haake1987classical,kus1987,Haake}. The latter term is responsible for the chaos in the system. Feingold and Peres (FP) introduced a time-independent coupled top without the nonlinear torsion term in the individual top \cite{FP,FP2}. However, the coupling between the two linear tops makes the system nonintegrable, and hence it shows chaos in some system parameter regimes. When the FP model was proposed, only three {\it standard} symmetry classes of Wigner-Dyson were known for the quantum systems \cite{wigner1951statistical,wigner1958distribution,dyson1962statistical,dyson1962threefold}. A recent study \cite{Non-Standard} has shown that the FP model follows one of the tenfold nonstandard symmetry classes \cite{verbaarschot1993spectral,verbaarschot1994spectrum,gade1993anderson,slevin1993new,altland1996random,Altland,zirnbauer1996riemannian}. The nonstandard symmetry classes were proposed following Cartan's tenfold symmetry space classification \cite{zirnbauer1996riemannian,zirnbauer2010symmetry}. Before the recent work in Ref. \cite{Non-Standard}, the implication of the nonstandard symmetry classes was not studied in the FP model.

This paper studies the FP model, which consists of two coupled identical linear tops. This model has permutation symmetry; additionally, it has chiral symmetry. The chiral symmetry in the FP model makes it a member of one of the nonstandard symmetry classes \cite{Non-Standard}. Moreover, we study a coupled top model with a nonlinear torsion term in each top. Depending on the symmetry class, the coupled top model with nonzero torsion (NZT) can further be classified into two models: NZT-I and NZT-II. In the case of the NZT-I model, the sign of the nonlinear torsion term of each top is the same. This model still has the permutation symmetry, but it loses chiral symmetry. Therefore, this model is no longer a member of any nonstandard symmetry classes; instead, it follows one of the standard three-fold Wigner-Dyson symmetry classes. The NZT-II model is not symmetric under permutation due to the torsion terms with different signs. However, like the FP model, this model also has chiral symmetry. Hence, the NZT-II model is again a member of one of the nonstandard symmetry classes.     

This paper focuses on classical and quantum mechanics of the FP and the NZT models. On the classical side, we study the amount of chaos in the systems by calculating their Lyapunov exponents. The effect of the underlying classical mechanics of all the models on their respective quantum versions is studied by calculating the entanglement between the two coupled tops and tracking the phase transition in the systems by varying the coupling strength between them. The study of the effect of classical mechanics on quantum entanglement is interesting because the latter is a pure quantum mechanical correlation without having any classical analogue \cite{schrodinger1935, einstein1935can,bells}. This pure quantum correlation remains intact even if one spatially separates two entangled parts of a system. This strange property of entanglement is used as a resource for most of the quantum information and computational protocols \cite{nielsen2002}. The remaining part of the paper concentrates on the phase transition in the FP model and the NZT-I and NZT-II coupled top models by tracking the transitions in the ground and the most excited states as a function of the coupling strength between the tops. These phase transitions can be attributed to the underlying classical dynamical transitions, like the bifurcation of the classical phase-space trajectories \cite{mondal2022quantum, mondal2022classical}. Overall, this paper explores the interplay of the underlying classical dynamics, entanglement, and phase transition in the three coupled top models (i.e., FP, NZT-I, and NZT-II models).   

We organize this paper as follows: Sec. \ref{model} introduces the coupled top model. Sections \ref{Class-Dyn} and \ref{Quant-Mech} discuss the classical and quantum mechanics of the system. The entanglement between the two tops in the coupled top model is studied in Sec. \ref{Entanglement}. In Sec. \ref{phase transition}, transitions in energies of the systems are presented. The paper is concluded in Sec. \ref{Conclusion}.
 
\section{Model}
\label{model}

We consider a coupled top model whose Hamiltonian is  
\begin{equation}
H = \Omega_1 J_{x1} +  \frac{\kappa_1}{2j} J_{z1}^2 + \Omega_2 J_{x2} + \frac{\kappa_2}{2j} J_{z2}^2 + \frac{\varepsilon}{j} J_{z1} J_{z2}.
\label{eq1}
\end{equation}
The first two terms represent the Hamiltonian of the individual top \cite{Haake}, and the third is the coupling between the two tops. This Hamiltonian is represented in terms of the angular momentum operators $\textbf{J}=(J_{xi},J_{yi},J_{zi})$ which follow usual angular momentum algebra $[J_{\alpha i}, J_{\beta i}] = i \epsilon_{{\alpha\beta\gamma},i}\, J_{\gamma i}$ and $i = 1, 2$ denotes individual top. The standard Levi-Civita symbol $\epsilon$ indicates the cyclic permutation in the commutator relations. As usual, the angular momentum operators corresponding to different tops commute, i.e., $[J_{\alpha 1}, J_{\beta 2}] = 0$, for $\alpha, \beta \in (x, y, z)$. We set the spin size of both the tops equal to $j$. The first term in the Hamiltonian describes the free precession of the individual top around $x$-axis with angular velocity $\Omega_i$, and the second term describes torsion in the individual top about $z$-axis by an angle proportional to $J_{zi}$, where the proportionality constant $\kappa_i$ is a dimensionless quantity. The third term describes the spin-spin coupling between the two tops of strength $\varepsilon$. 

The classical Hamiltonian $H_{\rm{cl}}$ is obtained from Eq. \eqref{eq1} by dividing the quantum Hamiltonian by spin $j$ and set the limit $j \rightarrow \infty$. At the large spin limit, the rescaled angular momentum operators $J_{\alpha i}/j$ commute with each other, and we obtain the classical Hamiltonian as:
\begin{equation}
H_{\rm{cl}} = \Omega_1 X_1 + \frac{1}{2} \kappa_1 Z_1^2 + \Omega_2 X_2 + \frac{1}{2} \kappa_2 Z_2^2 + \varepsilon Z_1 Z_2.
\label{H_cl}
\end{equation}
Here, $\alpha_i = J_{\alpha i}/j$, where $\alpha \in (X, Y, Z)$ denote different componets of the angular momentum and $i = 1, 2$ represents first and second tops. 

\section{Classical dynamics of the coupled top Hamiltonian}
\label{Class-Dyn}

From the classical Hamiltonian given in Eq. \eqref{H_cl}, we \textcolor{cyan}{can} derive classical equations of motion of the coupled top system using the generalized Poisson brackets introduced in Ref. \cite{Martin}. The same equations of motion can also be obtained as follows: first, we write Heisenberg's equation for each component of the angular momentum operator, i.e., $\frac{d J_{\alpha i}}{dt} = - i \left[J_{\alpha i}, H \right]$. Then divide both sides of the equation by the spin $j$. Finally, we use the commutation property of the rescaled angular momentum operators and obtain the equations of motion. The straightforwardness of the latter method prompts us to apply it to derive the classical equation of motion, and the following equations of motion are obtained (a detailed derivation is given in Appendix A)
\begin{align}
\dot{X}_1 &= - Y_1 \bigl(\kappa_1\, Z_1 + \varepsilon\, Z_2 \bigr), \nonumber\\
\dot{Y}_1 &= - \Omega_1\, Z_1 + X_1 \bigl( \kappa_1\, Z_1 + \varepsilon\, Z_2 \bigr),\nonumber\\
\dot{Z}_1 &= \Omega_1\, Y_1,\nonumber\\
\dot{X}_2 &= - Y_2 \bigl(\kappa_2\, Z_2 + \varepsilon\, Z_1 \bigr),\nonumber\\
\dot{Y}_2 &= - \Omega_2\, Z_2 + X_2 \bigl( \kappa_2\, Z_2 + \varepsilon\, Z_1 \bigr),\nonumber\\
\dot{Z}_2 &= \Omega_2\, Y_2.
\end{align}
The rescaled angular momentum variables satisfy the constraint $X_1^2 + Y_1^2 + Z_1^2 = X_2^2 + Y_2^2 + Z_2^2 = 1$. This suggests that the classical dynamics of the coupled top is restricted to the surface of $2$-spheres of unit radius. We can also exploit these constraints to reduce the degrees of freedom from {\it six} to {\it four} by transforming the Cartesian coordinates into spherical polar coordinates with unit radius. That is, we set the angular momentum variables as $X_i = \sin\theta_i\, \cos\phi_i,\,Y_i = \sin\theta_i\, \sin\phi_i,$ and $Z_i = \cos\theta_i$, where $Z_i$ and $\phi_i$ become canonically conjugate variables $\{Z_i, \phi_i\}=1$ for the $i$-th top. In terms of these new variables $(\theta_i, \phi_i)$, the classical Hamiltonian becomes
\begin{align}
H_{\rm cl} &= \Omega_1\, \sqrt{1-Z_1^2}\, \cos\phi_1 + \Omega_2\, \sqrt{1-Z_2^2}\, \cos\phi_2 \nonumber \\ &+ \frac{1}{2}\bigl(\kappa_1 Z_1^2 + \kappa_2 Z_2^2\bigr) + \varepsilon\, Z_1 Z_2,
\end{align}
and the corresponding Hamilton's equations of motion become:
\begin{align}
\dot{Z_1} &= -\frac{\partial H^{\rm cl}}{\partial \phi_1} = \Omega_1\,\sqrt{1-Z_1^2}\, \sin\phi_1 \nonumber\\
\nonumber\\
\dot{\phi_1} &= \frac{\partial H^{\rm cl}}{\partial Z_1} = \kappa_1\, Z_1 -  \frac{\Omega_1  Z_1}{\sqrt{1-Z_1^2}} \cos\phi_1 + \varepsilon\, Z_2,\nonumber\\
\nonumber\\
\dot{Z_2} &= -\frac{\partial H^{\rm cl}}{\partial \phi_2} = \Omega_2\,\sqrt{1-Z_2^2}\, \sin\phi_2\nonumber\\
\nonumber\\
\dot{\phi_2} &= \frac{\partial H^{\rm cl}}{\partial Z_2} = \kappa_2\, Z_2 - \frac{\Omega_2 Z_2}{\sqrt{1-Z_2^2}} \cos\phi_2 + \varepsilon\, Z_1.
\label{eq13}
\end{align}
For the torsion-free FP model, we set $\kappa_1 = \kappa_2 = 0$ in the above set of equations. On the other hand, for the NZT-I model, we consider $\kappa_1 = \kappa_2$; and for the NZT-II model, $\kappa_1 = -\kappa_2$, where $\kappa_i$ with $i=1,2$ denote the strength of the torsional term of the $i$-th top. The symmetry properties of these tops are discussed extensively in Sec. \ref{Quant-Mech}.

\subsection{FP model (torsion-free case)}

\begin{figure}
	\centering
\includegraphics[height=4cm,width=8cm]{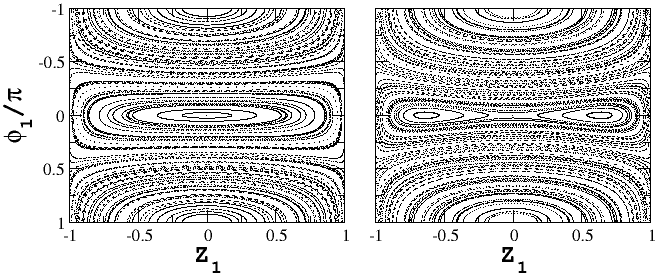}
\caption{The dynamics of the coupled top is projected on the phase space of the first top at the FP limit. The FP limit is obtained by setting $\kappa_1 = \kappa_2 = 0$. We set $\Omega_1 = \Omega_2 = 1.0$. Left panel: Coupling $\varepsilon = 0.8$; and Right panel: $\varepsilon = 1.3$.}
\label{Fig1}
\end{figure}

Figure \ref{Fig1} presents the phase space dynamics of the coupled top at the FP limit. Here, we fix $\Omega_1 = \Omega_2 = 1.0$ and consider two different coupling strengths $\varepsilon = 0.8$ (Left panel) and $\varepsilon = 1.3$ (Right panel). We have chosen these two coupling strengths because, at $\varepsilon = 1.2$, the phase space dynamics of the FP system show a transition, which is clear from the appearance of substructures at the center.

\subsection{Nonzero torsions case}

\begin{figure}
\centering
\includegraphics[height = 7cm, width=8cm]{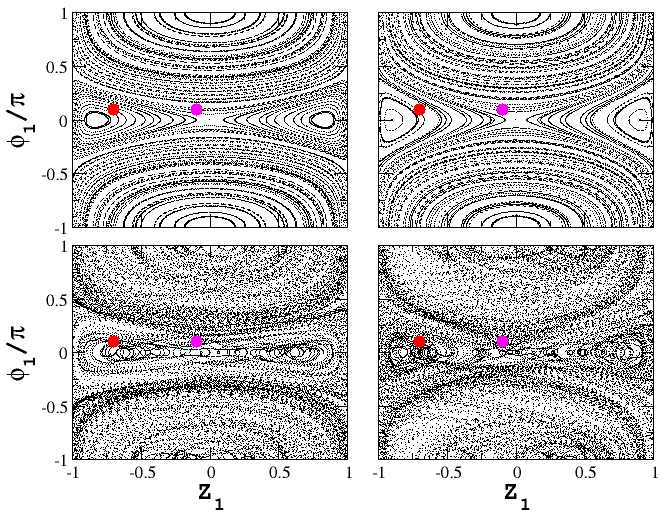}
\caption{The dynamics of the coupled top are projected on the phase space of the first top. We again set $\Omega_1 = \Omega_2 = 1.0$ and consider the same coupling strengths. Here, we have considered nonzero values of the torsion parameters $\kappa_1$ and $\kappa_2$. Top-Left panel: $\varepsilon = 0.8$ and $\kappa_1 = \kappa_2 = 1.0$; Top-Right panel: $\varepsilon = 1.3$ and $\kappa_1 = \kappa_2 = 1.0$; Bottom-Left panel: $\varepsilon = 0.8$ and $\kappa_1 = - \kappa_2 = 1.0$; Bottom-Right panel: $\varepsilon = 1.3$ and $\kappa_1 = - \kappa_2 = 1.0$. We observe completely different behavior of the phase space dynamics when we consider $\kappa_1$ and $\kappa_2$ of the same magnitude but of opposite signs.}
\label{Fig2}
\end{figure}

We present the classical dynamics of the Hamiltonian $H_{\rm cl}$  of the coupled top. Here again, we set the parameters $\Omega_1 = \Omega_2 = 1.0$ and consider the same two coupling cases, $\varepsilon = 0.8$ and $1.3$, to show the effect of the nonlinear torsional terms. In Fig. \ref{Fig2}, we consider two cases of torsions: at the top panels, two torsional terms are equal in magnitude and sign, i.e., $\kappa_1 = \kappa_2$, and named this NZT-I model. At the bottom panels, we consider torsional terms with opposite signs, but of the same magnitude, i.e., $\kappa_1 = - \kappa_2$, and named this NZT-II model. The phase space dynamics are qualitatively similar for the NZT-I model, presented in the top panels for both coupling strengths. This behavior is markedly different from the FP case, as discussed above. However, for the NZT-II model, the dynamics are very different from the case with the same value of the torsions. Here, we see that the trajectories are diffusing from one stable island to another at the center of the phase space of the first top. Here, we have not plotted the Poincar\'e section; instead, we have just projected the phase space trajectories of the coupled top residing on the four-dimensional phase space to the phase space of the first top having dimension two. Therefore, this diffusion of trajectories from one island to another happens through a higher-dimensional phase space. We also see the same behavior in other parts of the phase space.

\subsection{Lyapunov Exponent for nonzero torsions case}

\begin{figure}
\centering
\includegraphics[height=6cm,width=7cm]{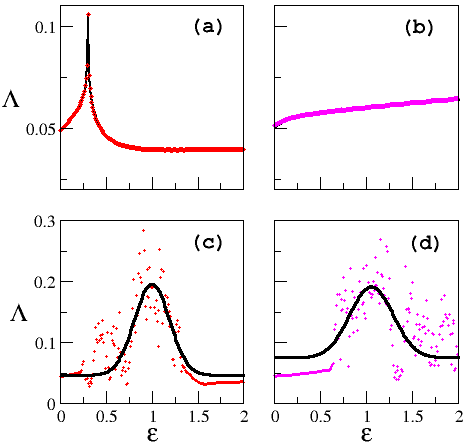}
\caption{Lyapunov exponent is plotted with the variation in coupling strength. Top row: For NZT-I ($\kappa_1 = \kappa_2 =1$) and Bottom row: For NZT-II ($\kappa_1 = - \kappa_2 =1$). Circles in the red and magenta colors in the figures show the Lyapunov exponent $\Lambda$, and the black solid line shows the fitted curve of the data.}
\label{LE}
\end{figure}

In this section, we have studied the Lyapunov exponent $\Lambda$ as a measure of chaos of the coupled top model. This measure is defined as 
\begin{equation}
\Lambda = \lim_{t\rightarrow\infty}\lim_{\Delta(0)\rightarrow0} \frac{1}{t}\ln\left[\frac{\Delta(t)}{\Delta(0)}\right].
\end{equation}
Here, \[\Delta(t) = \sqrt{\sum_{i=1}^2 \Delta X_i(t)^2 + \Delta Y_i(t)^2 + \Delta Z_i(t)^2}\] 
is the separation between two trajectories at time $t$, which were initially separated by a tiny amount $\Delta(0)\rightarrow0$. The Lyapunov exponent $\Lambda$ determines the exponential sensitivity of the initial condition of two neighboring trajectories. In Fig. \ref{LE}, we have shown $\Lambda$ as a function of the coupling strength $\varepsilon$ for the non-zero torsion cases of the model. We have considered two initial conditions: (i) $Z_1 = -0.7 $ and $ \Phi_1/ \pi = 0.1$ (shown as red circle in Fig. \ref{Fig2}), and (ii) $Z_1=-0.1$ and $\Phi_1/ \pi = 0.1$ (shown as meganta circle in Fig. \ref{Fig2}). 

In Figs. \ref{LE}(a) and (b), the results for NZT-I are presented for the above two initial conditions, whereas Figs. \ref{LE}(c) and (d) present the results for NZT-II for the same initial conditions. For the NZT-I model, $\Lambda$ increases with the coupling strength $\varepsilon$ for both the initial conditions. However, for the initial condition marked by a red circle in Fig. \ref{Fig2}, $\Lambda$ shows a peak $\varepsilon \simeq 0.3$; and then it goes down and saturates at a lower value when the coupling strength $\varepsilon \gtrsim 0.5$. For the other initial condition, we only observe a very slow increment in $\Lambda$ as a function of $\varepsilon$. In the NZT-II model, the Lyapunov exponent $\Lambda$ also shows variation with the coupling strengths $\varepsilon$. Moreover, the $\Lambda$ is greater than the NZT-I model for sufficiently strong coupling. This suggests that the NZT-II model is more chaotic. A transition in $\Lambda$ has been observed around $\varepsilon = 1.0$. We will see later that this transition in classical dynamics influences the entanglement transition and the transition in the excited state energy of the systems. This behavior of $\Lambda$ can be understood better from the behavior of a bunch of phase space trajectories with initial points within the magenta and red circles mentioned above. For the NZT-I and NZT-II models, the phase space trajectories are shown for different values of the coupling strengths $\varepsilon$ in Figs. \ref{Fig7} and \ref{Fig8} of Appendix \ref{appendixb}.

\section{Symmetry properties of the coupled top Hamiltonian}
\label{Quant-Mech}

\subsection{FP model (torsion-free case)}

For the FP model, the classical Hamiltonian of the coupled top becomes
\begin{equation}
H_{\rm FP} = \Omega_1 J_{x1} + \Omega_2 J_{x2}  + \frac{\varepsilon}{j} J_{z1} J_{z2}
\end{equation}
If $\Omega_1 = \Omega_2$, the above Hamiltonian has permutation symmetry, i.e., if we exchange the two tops $1 \leftrightarrow 2$, then the Hamiltonian remains invariant. Besides, the above Hamiltonian has a unitary symmetry 
\begin{equation}
U_0 = e^{-i \pi J_{x1}} \otimes e^{-i \pi J_{x2}},
\end{equation}
such that $U_0\, H_{\rm FP}\, U_0^\dagger = H_{\rm FP}$ or $\left[U_0, H_{\rm FP}\right] = 0$, where $U_0^2 = \mathbbm{1}$. The FP Hamiltonian has an additional unitary symmetry
\begin{equation}
C = e^{i \alpha} e^{-i \pi J_{z1}} \otimes e^{-i \pi J_{y2}},
\end{equation}
which gives $C\, H_{\rm FP}\, C^\dagger = -H_{\rm FP}$ or $\left(C\, H_{\rm FP} + H_{\rm FP}\, C\right) = 0$. This implies that ${\rm Tr}\,(H_{\rm FP}) = 0$. The phase factor $e^{i \alpha}$ is not playing any role here to show this symmetry of $H_{\rm FP}$ under $C$ because its complex conjugate factor from $C^\dagger$ part will trivially cancel it. However, this innocent-looking phase factor will become important while classifying the system under different symmetry classes. The operator, $C$, is called the chirality operator. In the case of the integer spin $C^2 = \mathbbm{1}$. This property of $H_{\rm FP}$ implies the presence of nonstandard symmetries different from the standard  Wigner-Dyson three-fold symmetry classes. In addition, the FP Hamiltonian has a nonunitary symmetry, like the time-reversal symmetry. We denote the time-reversal operator $\mathcal{T}$, which, on a standard basis, flips the sign of the $J_y$ operator but keeps the other two angular momenta invariant. Therefore, $H_{\rm FP}$ also satisfies $\mathcal{T}\, H_{\rm FP} \mathcal{T}^{-1} = H_{\rm FP}$. Depending on whether the chirality operator $C$ commutes with $\mathcal{T}$, we get different classes of symmetries for $H_{\rm FP}$. If the spin $j$ is integer, then we set $\alpha = 0$ and find that
\begin{align}
    \mathcal{T}\,C\, \mathcal{T}^{-1} &= \mathcal{T}\left[e^{-i \pi J_{z1}} \otimes e^{-i \pi J_{y2}}\right] \mathcal{T}^{-1}\nonumber\\
&= e^{i \pi J_{z1}} \otimes e^{-i \pi J_{y2}}\nonumber\\
&= e^{2 i \pi J_{z1}}\, e^{-i \pi J_{z1}} \otimes e^{-i \pi J_{y2}}\nonumber\\ &= C
\end{align}
where we use $\mathcal{T}\, i\, \mathcal{T}^{-1} = - i$; and for integer $j$, the term $e^{2 i \pi J_{z1}}$ becomes identity matrix. Thus we find $\mathcal{T}\,C\, \mathcal{T}^{-1} = C$. However, for the half-integer spin $e^{2 i \pi J_{z1}} \neq \mathbbm{1}$ and hence $\mathcal{T}\,C\, \mathcal{T}^{-1} \neq C$. For this case, the phase $\alpha$ can be tuned to get a $C$ which commutes with $\mathcal{T}$. For any arbitrary phase $\alpha$, we have,
\begin{align}
\mathcal{T}\, C\, \mathcal{T}^{-1} &= \mathcal{T}\left[ e^{i \alpha} e^{-i \pi J_{z1}} \otimes e^{-i \pi J_{y2}}\right] \mathcal{T}^{-1} \nonumber\\ 
&= e^{-i \alpha} e^{i \pi J_{z1}} \otimes e^{-i \pi J_{y2}} \nonumber\\
&= e^{- i 2 \alpha}\, e^{i 2 \pi J_{z1}}\, \left[ e^{i \alpha}\, e^{-i \pi J_{z1}}\, \otimes e^{-i \pi J_{y2}} \right] \nonumber\\
&=e^{i 2 (\pi J_{z1} - \alpha)}\, C.
\end{align}
If we set $\alpha = \pi\,j$, then for both integer and half-integer spin $j$, the first term of the last equality in the above equation will be an identity operator. Hence, we find that the chirality operator
\begin{equation}
C = e^{i \pi j} e^{-i \pi J_{z1}} \otimes e^{-i \pi J_{y2}}
\end{equation}
is time reversal symmetric, i.e., $\mathcal{T}\, C\, \mathcal{T}^{-1} = C$ and also transforms the FP Hamiltonian as $C\, H_{\rm FP} \, C^{-1} = - H_{\rm FP}$. We note that for the integer spin $j$, the chirality operator satisfies $C^2 = \mathbbm{1}$; and for the half-integer spin, $C^2 = - \mathbbm{1}$. Due to the presence of a time-reversal symmetric chirality operator, the FP model is classified as {\bf (i)} the BDI (BD One) class or the chiral orthogonal symmetry class for $C^2 = \mathbbm{1}$ (integer spin); and {\bf (ii)} the CI (C One) class or the anti-chiral class for $C^2 = - \mathbbm{1}$ (half-integer spin). These are two classes of the so-called nonstandard symmetries \cite{Altland, Non-Standard}. 

\subsection{Nonzero torsions case} 

For the nonzero torsions, first consider the NZT-I model ($\kappa_1 = \kappa_2 = \kappa$) and $\Omega_1 = \Omega_2 = \Omega$. Then the classical Hamiltonian $H_{\rm CT}$ of the coupled top becomes:
\begin{align}
H_{\rm CT} &= \Omega\bigl(J_{x1} + J_{x2} \bigr) + \frac{\varepsilon}{j} J_{z1} J_{z2} + \frac{\kappa}{2j} \left(J_{z1}^2 + J_{z2}^2\right) \nonumber\\ &= H_{\rm FP} + \frac{\kappa}{2j} \left(J_{z1}^2 + J_{z2}^2\right) \nonumber\\ &\equiv H_{\rm FP} + H_{\rm NL},
\end{align}
where $H_{\rm NL}$ is the nonlinear torsion part. This Hamiltonian is symmetric under permutation and remains invariant under the unitary transformation $U_0$ defined earlier. However, this Hamiltonian does not show chiral symmetry under the transformation $C$, i.e., $C H_{\rm CT} C^\dagger \neq - H_{\rm CT}$. This is simply because $H_{\rm NL}$ does not have chiral symmetry, i.e., $C\,H_{\rm NL}\,C^{-1} \neq - H_{\rm NL}$. This property is also a consequence of {\it nonzero} trace of $H_{\rm NL}$.

We now find the condition for which the trace of the Hamiltonian will be zero. Starting with different torsion strengths $\kappa_1 \neq \kappa_2$, we calculate the trace of the Hamiltonian on the standard basis as:
\begin{align}
{\rm Tr}\left(H_{\rm NL}\right) &= \frac{1}{2j}\left(\kappa_1 \sum_{m_1 = -j}^j m_1^2 + \kappa_2 \sum_{m_2 = -j}^j m_2^2\right) \nonumber\\
&=\frac{1}{j} \sum_{m=1}^j m^2 \left(\kappa_1 + \kappa_2\right) \nonumber\\
&= \frac{1}{6} (j+1) (2j+1) \left(\kappa_1 + \kappa_2\right).
\end{align}
The above relation shows that the trace is zero when $\kappa_1 = -\kappa_2 = \kappa$. Therefore, the coupled top Hamiltonian with trace zero is of the form
\begin{equation}
H_{\rm CT} = \Omega\bigl(J_{x1} + J_{x2} \bigr) + \frac{\varepsilon}{j} J_{z1} J_{z2} + \frac{\kappa}{2j} \left(J_{z1}^2 - J_{z2}^2\right).
\end{equation}
We have observed very different classical dynamics for this particular case of torsion with opposite signs. However, the Hamiltonian $H_{\rm CT}$ still lacks chiral symmetry under the transformation of $C$. This result implies that there exists a different chirality operator, say $C^\prime$, under which $C^\prime\, H_{\rm CT}\, C^{\prime \dagger} = - H_{\rm CT}$. One point is to be noted that now the Hamiltonian $H_{\rm CT}$ is not symmetric under permutation. We exploit this fact and obtain $C^\prime = P C$, where $P$ is the permutation operator. One can now easily check that $C^\prime\,H_{\rm CT}\, C^{\prime \dagger} = P\, C\,H_{\rm CT}\, C^\dagger P^\dagger= - H_{\rm CT}$ or $\left(C^{\prime}\,H_{\rm CT} + H_{\rm CT}\,C^{\prime}\right) = 0$. Since the chirality operator $C$ is time-reversal symmetric, one can show that the chirality operator $C^\prime$ for the coupled top Hamiltonian is also time-reversal symmetric. Moreover, for $C^2 = \pm \mathbbm{1}$, the other chirality operator also satisfies the same property, i.e., $C^{\prime 2} = \pm \mathbbm{1}$. Therefore, the coupled top Hamiltonian can also be classified into two chiral symmetry classes, BDI and CI \cite{Altland, Non-Standard}. Besides, for the nonzero trace case, one can not find a chirality operator under which the coupled top Hamiltonian will show the symmetry property. Therefore, this Hamiltonian belongs to the standard symmetry class.

\section{Entanglement Calculation}
\label{Entanglement} 

We now study the entanglement property of the coupled-top system represented by the Hamiltonian $H_{\rm CT}$ for zero (FP case) and nonzero torsion cases (NZT-I and NZT-II models). We study the entanglement of the ground and the most excited states of the coupled top Hamiltonian. Since these are pure states, we use the von Neumann entropy $S_V$ of the reduced density matrix (RDM) corresponding to these states as a measure of entanglement. The RDM corresponding to one of the tops is obtained by taking the partial trace over the other. If $|\psi\rangle$ is a state of the coupled top, then the RDM corresponding to one of the tops will be $\rho_i = {\rm Tr}_{\overline{i}}\left(|\psi\rangle\langle\psi|\right)$, where $(i, \overline{i}) = 1, 2 \,(\overline{i} = 2, 1)$ represents two tops. Then the von Neumann entropy is defined as  follows:
\begin{equation}
S_v = -{\rm Tr}_1[\rho_1\, \ln \rho_1] = -{\rm Tr}_2[\rho_2\, \ln \rho_2] = -\sum_i \lambda_i \ln \lambda_i. 
\end{equation}
Here, $\rho_1$ and $\rho_2$ share the same nonzero eigenspectrum $\{\lambda_i\}$.
 
\begin{figure}
\centering
\begin{tabular}{c}
~~~~~{\small Ground State}~~~~~~~~~~~~{\small Excited State}\\
\includegraphics[height=7cm, width=7.5cm]{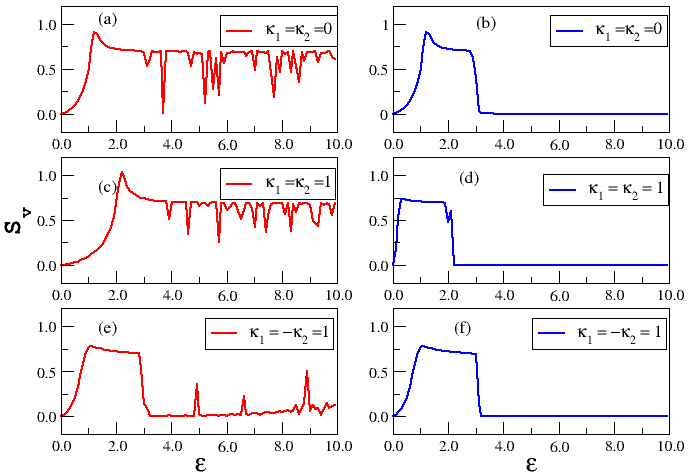}
\end{tabular}
\caption{Showing the variation of entanglement (von Neumann entropy) with coupling strength. The left column presents the results for the ground state, and the right column presents the results for the most excited state.}
\label{Fig3}
\end{figure}

Figure \ref{Fig3} presents the results of the entanglement calculation. Here, the variation of the von Neumann entropy is shown as a function of the coupling strength between the two tops $\varepsilon$ for the spin $j = 10$. We show the entanglement of two states: the ground state and the most excited state. Figure \ref{Fig3}(a) shows the result for the ground state energy of the FP model (i.e., $\kappa_1 = \kappa_2=0$). We find that, initially, $S_V$ increases linearly with $\varepsilon$ and attains a maximum value $S_V^{\rm max} \simeq 0.91$ at $\varepsilon = 1.2$, and it then immediately starts falling with the increment of the coupling strength. Later, we show that this transition in entanglement is consistent with the phase transition in the system. The entanglement falls till the coupling strength $\varepsilon \simeq 3.0$ and saturates at a value close to $\ln 2$ with some fluctuations. This saturation of $S_V$ around $\ln 2$ indicates permutation symmetry in the system, i.e., if we interchange the two tops, the Hamiltonian remains invariant. If we consider two non-identical coupled tops by setting unequal $\Omega$, the permutation symmetry will be broken, and consequently, the entanglement between the tops approaches zero for higher coupling strengths. In Fig. \ref{Fig3}(b), we present the results for the most excited state. Here also, the transition in the entanglement happens at the same coupling strength $\varepsilon = 1.2$. Interestingly, we do not see the effect of permutation symmetry on the excited state, and it becomes completely unentangled with $S_V = 0$ for stronger coupling strengths. 

In Figs. \ref{Fig3}(c) and (d), the von Neumann entropy $S_V$ is presented for the NZT-I model. In this case, like the previous one, the entanglement of the ground state initially increases linearly with the coupling strength $\varepsilon$ and reaches a maximum $S_V^{\rm max} \simeq 1.0$ at $\varepsilon = 2.0$. The entanglement then decreases and finally saturates at $\ln 2$ with some fluctuations for stronger coupling strengths $\varepsilon \gtrsim 4.0$. This saturation of entanglement at $S_V \simeq \ln 2$, also observed in the FP model, can be understood in the following way. At the strong coupling limit, the individual dynamics of the tops become almost negligible. The spin-spin coupling term in the Hamiltonian becomes the dominant term, and then the energy eigenstates approximately become $|m_1\rangle \otimes |m_2\rangle$, where  $|m_i\rangle$'s are the eigenstates of $J_{z_i}$ and $m_i = -j, \dots, +j$. These eigenstates are unentangled states. However, due to the presence of permutation symmetry in the system, $|\psi_{m_1, m_2}\rangle = \frac{1}{\sqrt{2}} \left(|m_1\rangle \otimes |m_2\rangle \pm |-m_1\rangle \otimes |-m_2\rangle\right)$ are also valid eigenstates with the same energy. The von Neumann entropy of these eigenstates $|\psi_{m_1, m_2}\rangle$ equals to $\ln 2$. However, this nonzero entanglement of these states is useless because it cannot be used as a resource for any quantum protocols. This is just an artifact of the underlying permutation symmetry in the system.

Figures \ref{Fig3}(e) and (f) show the results for the NZT-II model. This version of the coupled top system is interesting because it follows one of the nonstandard symmetry classes \cite{Altland, Non-Standard}. Since we consider integer spin here, the system follows the BDI symmetry class. In this case, the ground state entanglement also increases linearly and reaches its maximum $S_V^{\rm max} \simeq 0.8$ at $\varepsilon \simeq 1.0$ and suddenly starts decreasing. Here, the entanglement transition point is the same as the FP model. Like the previous two cases, the entanglement immediately goes down to $\sim \ln 2$ immediately after reaching the maximum, and then for $\varepsilon \simeq 3.0$, the system becomes unentangled. Interestingly, for $\varepsilon \gtrsim 3.0$, the von Neumann entropy increases slowly in a linear fashion with some fluctuations. The behavior of the excited state entanglement is almost similar to the ground state, except we observe that the excited state becomes unentangled or a product state at $\varepsilon \simeq 3.0$ and remains as such for larger values of coupling strength.

Overall, we find that the entanglement transition in the ground state of the FP model and the NZT-II model takes place at the same value of the coupling strength $\varepsilon \simeq 1.0$. However, the entanglement of the ground state at the transition point is higher for the FP model. Notably, the FP model has permutation symmetry, but the NZT-II model does not have that symmetry. Therefore, the saturation values of the entanglement for these two models are different: the FP model shows spurious nonzero entanglement with $S_V \sim \ln 2$, and the other model becomes completely unentangled and then increases slowly with the coupling strength. The entanglement of the most excited state behaves qualitatively similarly for both these models. Now, for the NZT-I model, the ground state entanglement makes a transition at a larger coupling strength $\varepsilon \simeq 2.0$ and then the entanglement reaches the saturation value $S_V = \ln 2$ around $\varepsilon \simeq 4.0$. The entanglement of the excited state of this model transitions at a very small value of coupling strength $\varepsilon = 0.3$, after that the entanglement decreases till $\varepsilon = 2.0$. The model transitions to the unentangled state in the $\varepsilon  >  2.0$ regime.

\section{Transition in energies and correlation with entanglement}
\label{phase transition}

We now investigate the phase transition in the system by studying the variation in the FP model's excited state and ground state energies and the coupled top models with non-zero torsion cases. We find that these transitions are related to the transition in the entanglement between the two tops. We analyze all the results from the underlying classical dynamics. Notably, we have shown how the system's stable-to-unstable transition of the classical fixed point (CFP) influences the quantum transitions. We observe that the quantum transitions and the transition in the classical fixed points (CFPs) stability occur at the same coupling strength $\varepsilon$ for torsion-free and nonzero torsional models. 
 
\begin{figure}[t]
\includegraphics[height=5cm, width=5.5cm]{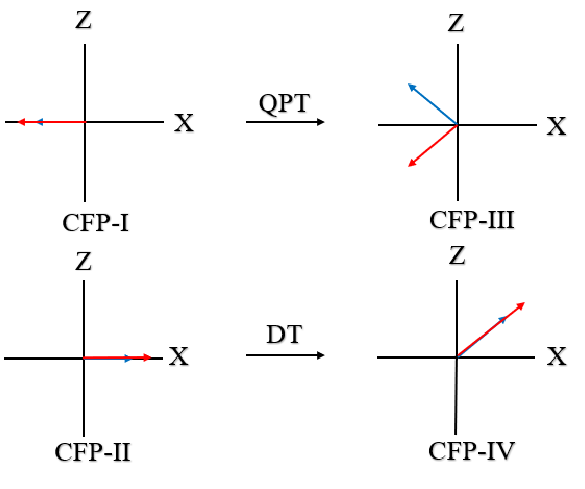}
\caption{Schematic diagram of the (classical) spin configuration for different steady states.}
\label{Fig4}
\end{figure}

The CFPs are determined by setting the time-derivatives of the dynamical variables $(Z_i, \phi_i)$ for $i = 1, 2$ at the left-hand side of Eq. \eqref{eq13} equal to zero. We now obtain a set of four coupled homogeneous algebraic equations, and the solutions of these equations give the CFPs of the system for different coupling strengths $\varepsilon$. We find the same four CFPs for the torsion-free and nonzero torsional cases: 
\begin{equation}
\begin{split}
\text{CFP-I}:&~ Z_1 = Z_2 = 0,\, \phi_1 = \phi_2 = \pi,\\ 
\text{CFP-II}:&~ Z_1 = Z_2 = 0,\,\phi_1 = \phi_2=0,\\ 
\text{CFP-III}:&~ Z_1 = -Z_2,\, \phi_1 = \phi_2 = \pi,\\ 
\text{CFP-IV}:&~ Z_1 = Z_2,\, \phi_1 = \phi_2 = 0.
\end{split}
\end{equation}
The generic behavior of the CFPs of the system is the following: Initially, for very weak coupling strengths $\varepsilon \gtrsim 0.0$, the system has two 
fixed points (CFP-I and CFP-II), when both the spins of the coupled top are in the same direction ($X$-direction) with zero magnetization along the $Z$-axis, as shown in Fig. \ref{Fig4}. This figure schematically shows that CFP-I and CFP-II are distinguished from the direction of the spin of the individual top: for CFP-I, both spins are along the negative $X$-direction, whereas for the CFP-II case, the spins are along the positive $X$-direction. These two fixed points correspond to symmetry-unbroken stable fixed points. At some critical values of the coupling strength $\varepsilon = \varepsilon_{\rm c}$, the classical fixed points become unstable and transition from the CFP-I to the CFP-III by bifurcation in energies. However, the critical value $\varepsilon_{\rm c}$ differs for CFP-I and CFP-II. The CFP-I state bifurcates and forms a pair of symmetry-broken anti-ferromagnetic states, CFP-III, at a critical point. On the other hand, the CFP-II state bifurcates at a different critical point and forms a ferromagnetic state, CFP-IV. These transitions are shown by a schematic diagram in Fig. \ref{Fig4}.

\begin{figure}[t]
	\centering
\includegraphics[height=7cm,width=8cm]{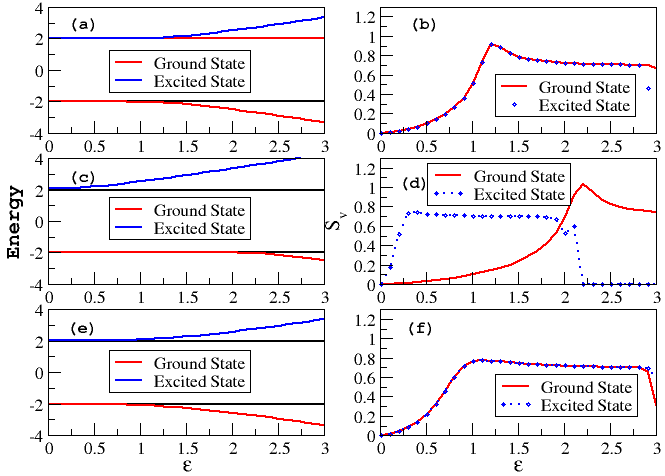}
\caption{Left column: Shows variation of energy (ground and most excited) as a function of coupling strength. Right column: Shows von Neumann entropy as a measure of entanglement with varying coupling strength.}
\label{Fig5}
\end{figure}

We now present a detailed analysis of the transitions in energies for the torsion-free and nonzero torsional cases of the coupled top. In the left column of Fig. \ref{Fig5}, we show the variation of the ground and the excited state energies with the coupling strength $\varepsilon$ to analyze the transition in energies of the system. In the right column, we have shown the corresponding variation of the quantum entanglement between the two tops. However, here we focus on the transition in energies and its relation with the entanglement, so we have shown the entanglement for the coupling strengths $\varepsilon \in [0.0, 3.0]$.

\subsection{FP model (torsion-free case)}

In the FP model, Fig. \ref{Fig5}(a) shows the ground and excited states energy variation with the coupling strength $\varepsilon$. Here, the bifurcation is observed in both states. The ground state is stable at $E = -2.0$ when $\varepsilon < 1.2$ (CFP-I state). Then, this state transitions to the CFP-III state at $\varepsilon = 1.2$, which remains stable for $\varepsilon > 1.2$. In the CFP-III state, the fixed point varies with the coupling strength as the ground state energy varies with the coupling strength as $Z_1 = - Z_2 = \sqrt{1- \frac{1}{\varepsilon^2}}$ with $\phi_1 = \phi_2 = \pi$, and consequently, the ground state energy varies as $E = -(\varepsilon+\frac{1}{\varepsilon})$. 

On the other hand, for $\varepsilon < 1.2$, the excited state energy of the system is stable at $E = +2$ (CFP-II state). This state transitions to the CFP-IV state by bifurcation at the critical point $\varepsilon = 1.2$, and then it remains stable for $\varepsilon > 1.2$. After the transition of the excited state from CFP-II to CFP-IV state, the fixed point also varies with $\varepsilon$ in the same fashion as the CFP-III state of the ground state. However, the excited state energy now increases as $E = (\varepsilon+\frac{1}{\varepsilon})$.      

For the FP model, we observe that the transitions in ground and excited states are happening around the same value of the coupling strength $\varepsilon \simeq 1.0$. Figure \ref{Fig5}(b) shows the variation in entanglement (von Neumann entropy) of the ground state (in red color) and excited state (in blue color) as a function of the coupling strength. We find that the entanglement of the ground state and the excited state maintains a steady growth till and then suddenly falls at $\varepsilon \simeq 1.0$. Interestingly, this transition in the entanglement is concurrent with a transition in energies and the stability of the fixed points of the underlying classical dynamical system.  

\subsection{Nonzero torsion case}

We now present the results for two nonzero torsional coupled top models: (a) NZT-I model ($\kappa_1 = \kappa_2 = 1.0$) and (b) NZT-II model ($\kappa_1 = - \kappa_2 = 1.0$).
 
\subsubsection{NZT-I model ($\kappa_1 = \kappa_2 = 1.0$)}

Figure \ref{Fig5}(c) shows the variation in ground state and excited state energy as a function of the coupling strength $\varepsilon$. This figure shows that the CFP-I state becomes unstable at $\varepsilon = (\kappa_1 + 1) = 2$ and bifurcates into the CFP-III state. The energy of the CFP-III state varies with the coupling strength as  $E = (1-\varepsilon) + \frac{1}{1-\varepsilon}$. On the other hand, the CFP-II state becomes unstable when $\varepsilon > (-\kappa_1 +1) = 0$. However, the effect of the unstable CFP-II state becomes prominent at a finite but very small value $\varepsilon \simeq 0.23$, where the CFP-II state bifurcates into the CFP-IV state. The energy of the CFP-IV state follows $E = (1+\varepsilon)+\frac{1}{(1+\varepsilon)}$. In Fig. \ref{Fig5}(d), the transition in entanglement is again exactly happening at the point of bifurcation for the ground and excited states. However, unlike the FP model, the transitions in the ground and excited states do not coincide at the same coupling strength. Still, these transitions are consistent with the corresponding entanglement transition and the bifurcation of the steady states.  

\subsubsection{NZT-II model ($\kappa_1 = - \kappa_2 = 1.0$)} 

The results corresponding to this case are shown in Figs. \ref{Fig5}(e) and (f). From Fig. \ref{Fig5}(e), we observe the transitions in the ground and excited states simultaneously at $\varepsilon \simeq 1.0$. This result is similar to the FP model case. Here, at the classical limit, the effect of torsion is canceled because of the opposite sign with the same magnitude of the torsion strength of each top. The energy for the CFP-III is calculated as $E = -\frac{2}{1-\varepsilon}-\varepsilon\left[1-\frac{1}{(1-\varepsilon)^2}\right]$; whereas for the CFP-IV state, the energy follows the relation $E = \frac{2}{1+\varepsilon}-\varepsilon\left[1-\frac{1}{(1+\varepsilon)^2}\right]$. Figure \ref{Fig5}(f) shows the transition in entanglement, which is consistent with the energy transition.

The above analysis reveals that the energy transition coincides around $\varepsilon \simeq 1.0$ in the FP and NZT-II models. A common feature of these two models is that they both have chiral symmetry and thus follow one of the nonstandard symmetry classes. Since we only consider integer spin here, these models follow the BDI symmetry class. On the other hand, the NZT-I model does not follow chiral symmetry. Interestingly, the transition in energies does not coincide in this model. However, these transitions are consistent with the corresponding entanglement transition and the transition in the stability of the fixed point of the underlying classical system.

\section{Conclusion}
\label{Conclusion}

We have studied the classical and quantum properties of the coupled top model in the absence and the presence of nonlinear torsion in the individual top. In the absence of nonlinear torsion of each top, the system becomes the well-known Feingold-Peres (FP) model. Besides the FP model, we also extensively study the coupled top model for two nonzero torsion (NZT) cases: torsion strengths of the individual top are equal (NZT-I), and torsion strengths are equal in magnitude but opposite in sign (NZT-II). The quantum version of the FP model and the NZT-II model follow the BDI symmetry class or chiral orthogonal symmetry class, one of the nonstandard symmetry classes. However, these two models are different from the perspective of permutation symmetry: the FP model has this symmetry, but the NZT-II model does not. On the other hand, the NZT-I model has permutation symmetry but does not have chiral symmetry. Therefore, the NZT-I model does not follow any nonstandard symmetry class; it is a member of one of the standard threefold symmetry classes of Wigner-Dyson. Interestingly, we have obtained the NZT-II model from the NZT-I model by breaking the permutation symmetry via setting the torsion strengths of the two tops opposite in sign. The breaking of the permutation symmetry in the NZT-I model facilitates the construction of a chirality operator for the NZT-II model. We have also studied entanglement between the two tops as a function of the coupling strength by calculating the von Neumann entropy. The most important outcome of this study is that, for the systems with BDI symmetry class (FP and NZT-II models),  
the transition in ground and excited energies and the transition in the entanglement happen simultaneously. However, in the NZT-I model, 
the transitions in the same energy states do not coincide; however, individually, these transitions coincide with the corresponding entanglement transition. Interestingly, all these transitions are consistent with the stability of the fixed points of the underlying classical dynamical systems.

\appendix

\section{Equations of Motion}

Following the prescription defined in Sec. \ref{Class-Dyn}, we present a rigorous derivation of the equation of motion of only one component of the angular momentum, say $J_{x1}$. Others are derived identically, and we present the final form of them. Equation of motion of angular momentum operator ${J}_{x1}$ can be obtained as:
\begin{align}
\frac{d {J}_{x1}}{dt}&=-i [{J}_{x1},{H}_{\rm eff}]\nonumber\\ 
&=-\frac{i}{2j} \biggl( \kappa_1\, [{J}_{x1},{J}_{z1}^2] + 2 \varepsilon\, [{J}_{x1},{J}_{z1}] {J}_{z2}\biggr)\nonumber\\
&=-\frac{i}{2j}\biggl\{\kappa_1\,\bigl([{J}_{x1},{J}_{z1}]{J}_{z1}+ {J}_{z1}[{J}_{x1},{J}_{z1}]\bigr)\nonumber\\ &~~~~~~+ 2\varepsilon\, [{J}_{x1},{J}_{z1}]\, {J}_{z2}\biggr\}\nonumber\\
&=-\frac{1}{2j}\biggl(\kappa_1 \bigl({J}_{y1} {J}_{z1}+{J}_{z1} {J}_{y1}\bigr)+ 2 \varepsilon\, {J}_{y1} {J}_{z2}\biggr).
\end{align}
Now, divide the above equation by $j$ and obtain the form:
\begin{equation}
\frac{1}{j} \frac{d J_{x1}}{dt} = - \frac{\kappa_1}{2} \left(\frac{J_{y1}}{j} \frac{J_{z1}}{j} + \frac{J_{z1}}{j} \frac{J_{y1}}{j}\right) - \varepsilon\,\frac{J_{y1}}{j} \frac{J_{z2}}{j}.
\end{equation}
We now set $j \rightarrow \infty$ limit, and get the classical equation of motion as:
\begin{equation}
\frac{d X_1}{dt} = - Y_1 \bigl(\kappa_1\, Z_1 + \varepsilon\, Z_2 \bigr).
\label{X1}
\end{equation}
Following the above steps, we obtain the full equation of motion:
\begin{align}
\dot{X}_1 &= - Y_1 \bigl(\kappa_1\, Z_1 + \varepsilon\, Z_2 \bigr), \nonumber\\
\dot{Y}_1 &= - \Omega_1\, Z_1 + X_1 \bigl( \kappa_1\, Z_1 + \varepsilon\, Z_2 \bigr),\nonumber\\
\dot{Z}_1 &= \Omega_1\, Y_1,\nonumber\\
\dot{X}_2 &= - Y_2 \bigl(\kappa_2\, Z_2 + \varepsilon\, Z_1 \bigr),\nonumber\\
\dot{Y}_2 &= - \Omega_2\, Z_2 + X_2 \bigl( \kappa_2\, Z_2 + \varepsilon\, Z_1 \bigr),\nonumber\\
\dot{Z}_2 &= \Omega_2\, Y_2.
\label{Class-EOM}
\end{align}

\section{Phase space}
\label{appendixb}

In this section, we have plotted the phase space dynamics of NZT-I and NZT-II models for different coupling strength $\varepsilon$, around the two initial conditions shown in red and magenta circles in Fig. \ref{Fig2}.
\begin{figure}
\includegraphics[height=7cm, width=8cm]{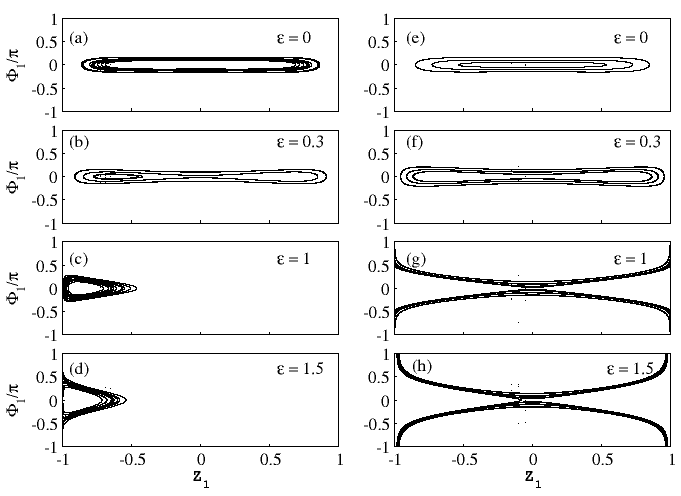}
\caption{A bunch of trajectories is shown for the NZT-I model. (a)-(d): The trajectories are initiated around $Z_1 = -0.7$ and $\Phi_1 / \pi= 0.1 $ shown by a solid red circle in Fig. \ref{Fig2}. (e)-(h): The trajectories are initiated around $Z_1 = -0.1$ and $\Phi_1/\pi = 0.1 $, shown by a solid magenta circle in Fig. \ref{Fig2}.}
\label{Fig7}
\end{figure}
In Figs. \ref{Fig7}(a)-(d), a bunch of trajectories are plotted for the initial conditions around $Z_1 = -0.7$ and $\Phi_1/\pi = 0.1$ for different coupling strengths. On the other hand, Figs. \ref{Fig7}(e)-(h) show the trajectories initiated around another point $Z_1 = -0.1$ and $\Phi_1/\pi = 0.1$. Figures \ref{Fig7}(a) and (e) suggest that, for the uncoupled case $\varepsilon = 0$, identical regular islands are visible for both the initial conditions. 

As the coupling strength increases $\varepsilon = 0.3$, Fig. \ref{Fig7}(b) indicates the beginning of transitions in the phase space. This figure shows that the observed single regular region for the uncoupled case is now bifurcating and forming two separate regular islands. This transition was detected in the Lyapunov exponent result presented in Fig. \ref{LE}(a). However, we do not observe any bifurcation of the regular island for the other initial condition shown in Fig. \ref{Fig7}(f). The corresponding Lyapunov exponent calculation in Fig. \ref{LE}(b) also does not show any peaks, which is consistent with the behavior of the phase space trajectories.   

We further increase the coupling strength $\varepsilon = 1.0$. Figure \ref{Fig7}(c) shows complete bifurcation of the initial single regular region into two regular regions. However, this figure shows one regular island at the left, because we have only considered the trajectories initiated around the red solid circle shown in Fig. \ref{Fig2}. In \ref{Fig7}(d), we do not see any further transition in the phase space trajectories for stronger coupling strength, except for a minute movement of the regular island. Therefore, the absence of any peaks in the Lyapunov exponent result in Fig. \ref{LE}(a) for stronger coupling strengths is consistent with the behavior of the phase space trajectories. For the other initial conditions, the phase-space trajectories initiated from the region marked by the solid magenta circle in Fig. \ref{Fig2} show significantly different behavior. However, since the phase-space has the periodic property (i.e., left and right boundaries are the same line, whereas top and bottom boundaries are also the same line), a careful observation of Figs. 7(g) and (h) will reveal that each initial condition generates a closed loop on the surface of the sphere, and consequently, the trajectories are not sensitive to the initial conditions. This property is also reflected in the Lyapunov exponent result presented in Fig. \ref{LE}(b). 

\begin{figure}
\includegraphics[height=7cm, width=8cm]{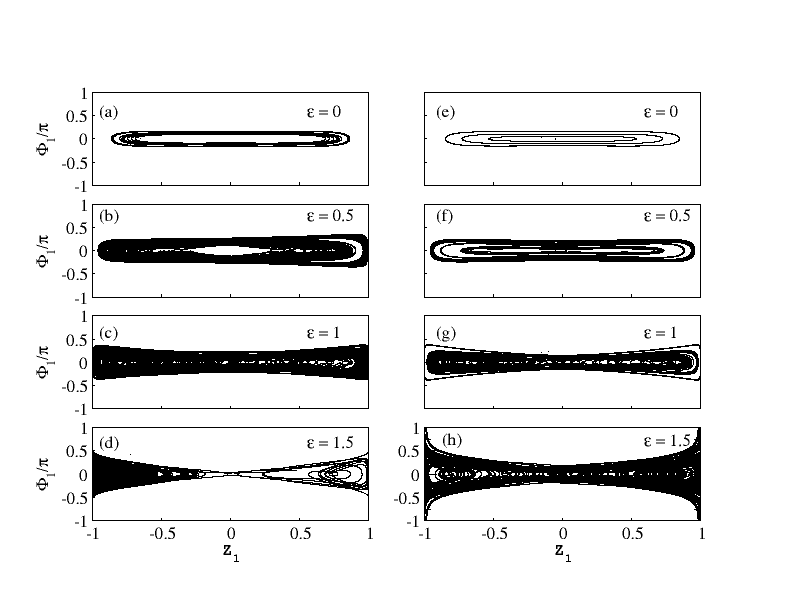}
\caption{A bunch of trajectories are shown for the NZT-II model. (a)-(d): The trajectories are initiated around $Z_1 = -0.7$ and $\Phi_1/\pi = 0.1$ shown by a solid red circle in Fig. \ref{Fig2}. (e)-(h): The trajectories are initiated around $Z_1 = -0.1$ and $\Phi_1/\pi = 0.1$, shown by a solid magenta circle in Fig. \ref{Fig2}.}
\label{Fig8}
\end{figure}

We have also plotted many trajectories around the same two initial conditions for the NZT-II model in Fig. \ref{Fig8}. For $\varepsilon = 0$, regular islands are observed for both the initial conditions, as shown in Figs. \ref{Fig8}(a) and (e). For $\varepsilon = 0.5$, transitions in trajectories have been observed in both cases. In the NZT-II model, the bifurcation of the trajectories starts around $\varepsilon=0.5$. At $\varepsilon = 1.0$, we see two prominent regular islands within a regular envelope trajectory observed in the uncoupled tops. However, unlike the single top, the trajectories start from one island and can reach another by the Arnold diffusion process, a typical phenomenon observed in higher-dimensional nonintegrable systems. For higher values of coupling strength $\varepsilon > 1.5$, we do not see any qualitative change in the phase space trajectories. The corresponding Lyapunov exponent result shows the beginning of its steady growth at $\varepsilon = 0.5$, reaching the maxima at $\varepsilon = 1.0$, and then again goes down. Furthermore, we also conclude that the transition in phase space dynamics of NZT-I has been observed at $\varepsilon=0.3$, and for NZT-II it is observed at $\varepsilon=1.0$. These results are consistent with the observed transition in the entanglement between the two tops in the ground and most excited state energies.

\begin{acknowledgments}
JNB acknowledges the Department of Science and Technology (DST) for providing computational resources through the FIST program (Project No. SR/FST/PS-1/2017/30). JNB also acknowledges financial support from DST-SERB, India, through a MATRICS grant MTR/2022/000691.
\end{acknowledgments}

\bibliography{CKT}

\end{document}